\documentclass{IEEEtran}
\usepackage{cite}
\usepackage{amsmath,amssymb,amsfonts}
\usepackage{graphicx}
\usepackage{textcomp,nicefrac}
\usepackage{xcolor}
\usepackage{tabularx}% 导言区添加
\usepackage{booktabs}
\usepackage{url}
\usepackage[switch]{lineno}

\def\BibTeX{{\rm B\kern-.05em{\sc i\kern-.025em b}\kern-.08em
T\kern-.1667em\lower.7ex\hbox{E}\kern-.125emX}}
\begin{document}
%\linenumbers
\title{Performance Test and Circuit Simulation for R12699-406-M4 Photomultiplier Tube Base}

\author{Houqi~Huang, 
        Peiyuan~Chen, 
        Ke~Han, 
        Yang~Liu,
        Guanbo~Wang, 
        Shaobo~Wang, 
        Weihao~Wu, 
        Binbin~Yan,  
        Peihua~Ye, 
        Jiaxu~Zhou, 
        and Zhizhen~Zhou%
\thanks{
 
Houqi~Huang and Peiyuan~Chen contributed equally to this work. 
(Corresponding authors: Shaobo~Wang; Binbin~Yan.)}%

\thanks{Houqi~Huang, Shaobo~Wang, and Jiaxu~Zhou are with SJTU Paris Elite Institute of Technology (SPEIT), 
Shanghai Jiao Tong University, Shanghai 200240, China 
(e-mail:shaobo.wang@sjtu.edu.cn).}%

\thanks{Peiyuan~Chen is with the 
State Key Laboratory of Dark Matter Physics, Key Laboratory for Particle Astrophysics and Cosmology (MoE), 
Shanghai Key Laboratory for Particle Physics and Cosmology, School of Physics and Astronomy, 
Shanghai Jiao Tong University, Shanghai 200240, China, and enrolled at the School of Physics, Jilin University, Changchun 130012, China}%

\thanks{Ke~Han, Yang~Liu, Guanbo~Wang, Weihao~Wu, Peihua~Ye, and Zhizhen~Zhou are with the 
State Key Laboratory of Dark Matter Physics, Key Laboratory for Particle Astrophysics and Cosmology (MoE), 
Shanghai Key Laboratory for Particle Physics and Cosmology, School of Physics and Astronomy, 
Shanghai Jiao Tong University, Shanghai 200240, China.}%

\thanks{Binbin~Yan is with the State Key Laboratory of Dark Matter Physics, Key Laboratory for Particle Astrophysics 
and Cosmology (MoE), Shanghai Key Laboratory for Particle Physics and Cosmology, Tsung-Dao Lee Institute \& School of 
Physics and Astronomy, Shanghai Jiao Tong University, Shanghai 201210, China 
(e-mail: yanbinbin@sjtu.edu.cn).}%

\thanks{Ke~Han, Shaobo~Wang and Binbin~Yan are also with the Shanghai Jiao Tong University Sichuan Research Institute, 
Chengdu 610213, China.}%

\thanks{Ke~Han, Shaobo~Wang and Weihao~Wu are also with the 
Jinping Deep Underground Frontier Science and Dark Matter Key Laboratory of Sichuan Province, 
Liangshan 615000, China.}%
}

\maketitle

\begin{abstract}
The next-generation liquid xenon experiments like PandaX-xT target an energy range from sub-keV to multi-MeV to address the requirement of multiple physics searches. The Hamamatsu R12699-406-M4 photomultiplier tubes (PMTs) were developed and selected as photon sensors for PandaX-xT. Their voltage-divider base is optimized for a broad dynamic range, from single-photoelectron (SPE) sensitivity to 30~nC collected charge (matching the 2.5~MeV Q-value of $^{136}$Xe neutrinoless double beta decay~(NLDBD)). Using a dedicated test bench, we characterize the saturation and suppression responses of R12699-406-M4 PMTs with this base design. Based on measured PMT-base responses, we develop a circuit simulation model that accurately reproduces the physical mechanisms underlying these effects with key parameters tuned via experimental data. The combined simulation and bench-test approach guides base design and optimization, enabling improved detector dynamic range and supporting future saturation and suppression correction studies in data analysis.

\end{abstract}


\section{Introduction}

\label{sec:introduction}

\IEEEPARstart{L}{iquid} xenon~(LXe) detectors are one of the most sensitive techniques for dark matter (DM) searches~\cite{Baudis:2023pzu}. These searches primarily target weakly interacting massive particles (WIMPs)~\cite{XENON:2024wpa}, hypothesized particles proposed to explain dark matter. With increasing target mass, LXe detectors become promising for rare-event studies, including solar neutrinos~\cite{PandaX4T_CEνNS_2024} and NLDBD~\cite{PandaX:2024fed,PandaX:2023ggs}. 
Thus, next-generation LXe detectors need to achieve sensitivity across an energy range from sub-keV to several MeV, which is imperative for simultaneous searches for DM and NLDBD.

Currently, large-scale LXe experiments, such as PandaX-4T~\cite{PandaX:2024qfu}, LZ~\cite{LZ2025}, and XENONnT~\cite{XENONnT2025}, have demonstrated the capability to search for both WIMPs and NLDBD using the same detector. 
For example, PandaX-4T contains 3.7~tonnes of natural xenon in the sensitive volume, and has achieved a WIMP-nucleon cross-section limit of $1.6\times10^{-47}\mathrm{cm}^2$ at a WIMP mass of 40~GeV/$c^2$~\cite{PandaX:2024qfu}, a $^{136}$Xe NLDBD half-life limit of $T_{1/2}>2.1\times10^{24}$~yr~\cite{PandaX:2024fed}, and the most stringent half-life limits on $^{134}$Xe double beta decay to date~\cite{PandaX:2023ggs}. 
Next-generation detectors, including XLZD~\cite{Baudis:2024jnk}, and PandaX-xT~\cite{Abdukerim2025}, will incorporate multi-ten-tonne active targets and reduced backgrounds, enabling WIMP sensitivity to reach the neutrino floor while significantly enhancing NLDBD sensitivity.

LXe detectors utilize a dual-phase xenon time projection chamber (TPC)~\cite{Aalbers:2022dzr}, recording prompt scintillation (S1) and delayed electroluminescence (S2) signals simultaneously via top and bottom PMT arrays of the detector. 
This enables full three-dimensional position reconstruction and precise energy measurements of an event. 
The S1 signal, originating at the event vertex, generates relatively uniform light distributions across both PMT arrays. 
For the S2 signal, ionization electrons drift upwards to the liquid xenon surface, where they are extracted into the gaseous xenon (GXe) phase and undergo proportional electroluminescent amplification. 
Consequently, the top PMT array shows a concentrated hit pattern, with the brightest PMT typically collecting approximately one-third of the total charge. 
In PandaX-4T, the 3-inch R11410 PMTs in the top array suffered severe saturation effects at high energy range, degrading energy resolution and position reconstruction accuracy ~\cite{PandaX:2022kwg}.
We have achieved a linear response up to 32~nC at a gain of $5\times10^6$ by implementing upgraded PMT base circuits. 
To further extend the dynamic range up to the Q-value of $^{136}$Xe NLDBD (about 2.5~MeV), we developed a waveform desaturation method that aligns the rising edges of saturated waveforms with those of reference waveforms. 
This approach improves the linearity by a factor of 5~\cite{Luo2024}.

\begin{figure}[htbp]
\centerline{\includegraphics[width=\linewidth]{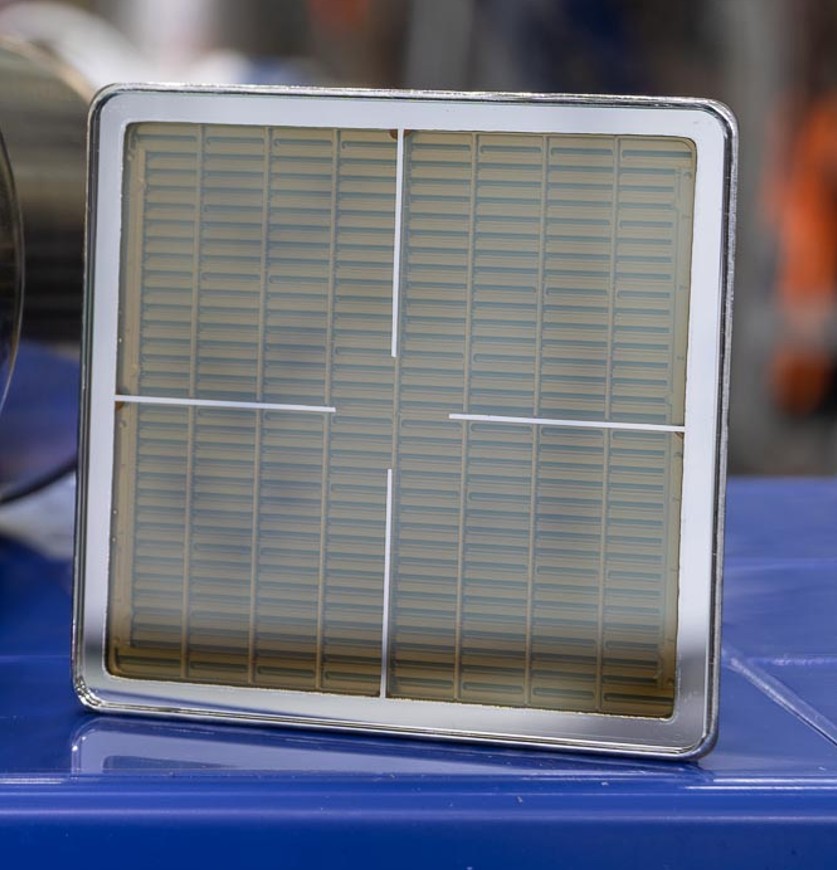}}
\caption{The picture of 2-inch R12699 PMT. The square PMT integrates four identical 1-inch detection
channels within a single housing.}
\label{fig:PMT}
\end{figure}

The PMT array is a critical detector component, sitting closest to the detector's sensitive volume and directly collecting signals. 
For the next-generation PandaX-xT detector, the R12699-406-M4 PMT~\cite{Yun:2024oxp}
%~\cite{Abdukerim2025, Yun:2024oxp}
(abbreviated R12699), was developed in collaboration with Hamamatsu Photonics K.K., featuring substantially reduced intrinsic radioactivity.
As shown in Fig.~\ref{fig:PMT}, this 2-inch square PMT integrates four identical 1-inch detection channels within a single housing, enhancing granularity. 
All four channels share a common dynode amplification system through the base circuit. Based on the performance of the 3-inch R11410 PMTs in PandaX-4T and scaled by the photocathode area ratio, the R12699 PMT requires a linearity range of 30 nC to cover energies up to 2.5 MeV at the same gain.

In this work, we systematically investigate the base configurations for the R12699 PMT using bench tests and circuit simulation.
Focusing on mitigating signal saturation and suppression effects, we propose an optimized design that meets the requirements of next-generation LXe detectors.
Section~\ref{design chapter} describes three versions of the PMT base with different numbers of capacitors.
In Section~\ref{test chapter}, we introduce a dedicated bench setup used to evaluate the saturation and suppression effects.
Section~\ref{simu chapter} presents the circuit-simulation framework for the PMT base and illustrates the mechanisms underlying saturation and suppression.
Finally, a summary is given in Section~\ref{sum chapter}.
 
\section{Base circuit design}
\label{design chapter}

\begin{figure*}[htbp]
\centerline{\includegraphics[width=\linewidth]{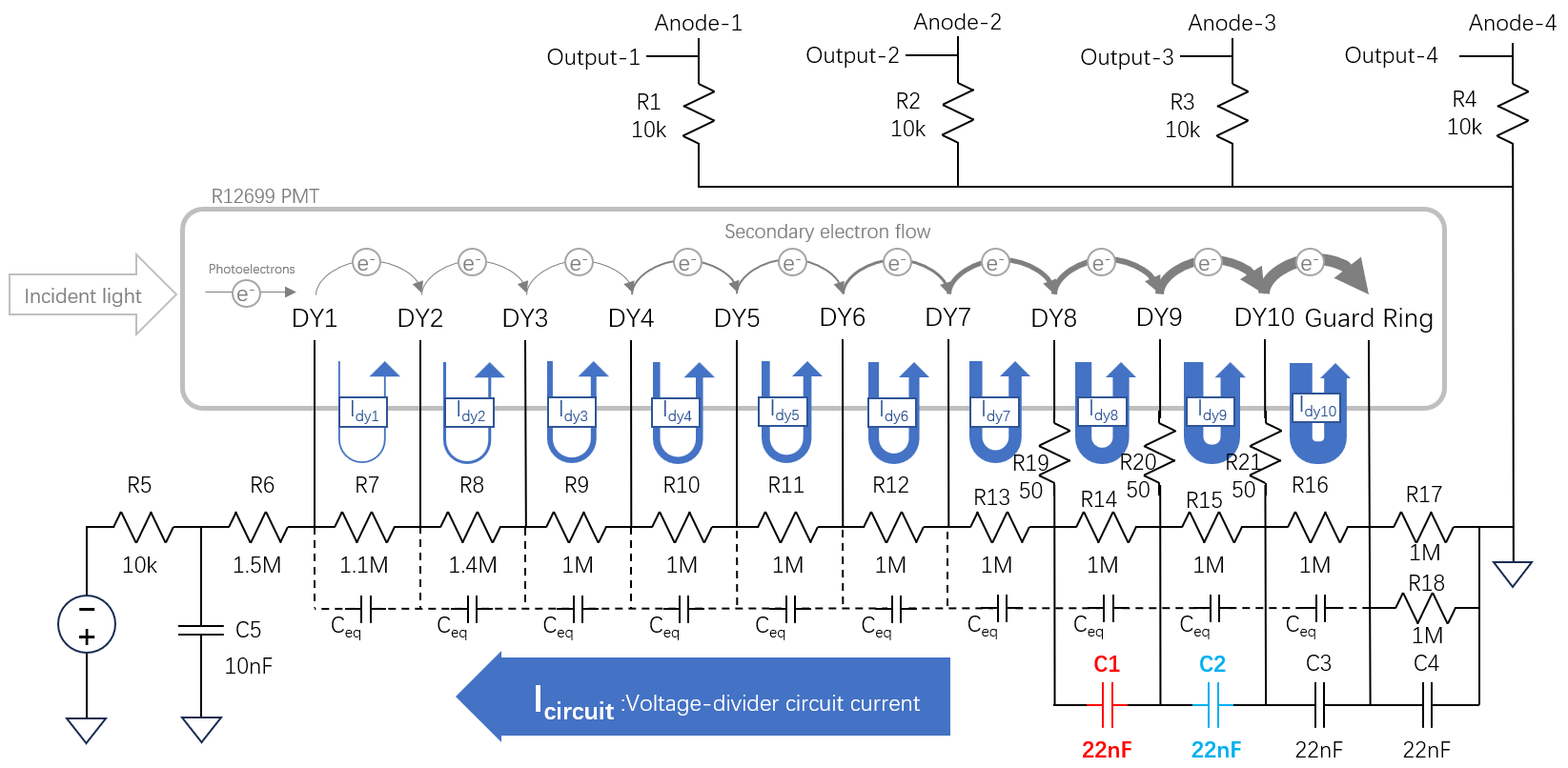}}
\caption{The designed base circuit of the R12699 PMT, illustrating the electron flow and currents during operation. Please refer the text for detailed information.}
\label{fig:base_design_v2}
\end{figure*}

The base denotes the voltage-divider circuit board, providing operating voltage division across the dynodes and enabling anode signal transmission.
As illustrated in Fig.~\ref{fig:base_design_v2}, the resistor chain ratio follows the base circuit design proposed by Hamamatsu~\cite{Hamamatsu2017}. The recommended operating voltage is 1000~V (photocathode-to-anode), with the anode grounded and $-$1000~V applied to the photocathode.

The resistance is selected as 1~M$\Omega$ by considering the thermal power tolerance in the LXe environment~\cite{Abdukerim2025}. 
The current through the voltage-divider circuit, $\mathrm{I}_\mathrm{circuit}$, maintains a constant current of approximately 80~{\textmu}A. Resistors R6--R18 establish an 80--112~V inter-stage voltage between adjacent dynodes. Three damping resistors (R19--R21) are placed at the last dynodes (Dy8-Dy10) to mitigate signal oscillation. 
Capacitors C1--C4 (22~nF each) suppress saturation effects, while C5 (10~nF) acts as a power-supply filter capacitor. 
The equivalent capacitance (C$_\mathrm{eq}$) at each dynode stage is 95.6$\pm$7.6~pF, measured via a Keysight E4980A LCR meter at 1~kHz (similar value at 300~kHz) and 25$^o$C~\cite{keysight}.

The multiplication factor per dynode stage ranges from 3 to 8 and increases with higher voltage division. Electron multiplication at each stage exponentially amplifies the electron flux, producing progressively larger secondary currents~($\mathrm{I}_{\mathrm{dy1}}$, $\mathrm{I}_{\mathrm{dy2}}$ ... $\mathrm{I}_{\mathrm{dy10}}$).
The secondary currents flow opposite to $\mathrm{I}_\mathrm{circuit}$, causing deviations from the inter-stage voltage distribution, and this type of deviation can be mitigated by desaturation capacitors.
The charge stored in C$_\mathrm{eq}$ and C1--C4 are approximately 8~nC and 1760~nC, respectively.
Under small-signal conditions, such deviations are negligible. 
For large signals, however, significant charge trapping occurs when the secondary charge induced by the signal is comparable to the charge stored at each dynode.
This leads to substantial voltage deviations that reduce the multiplication factors at later dynodes (typically from DY7 to DY10), thereby decreasing the output signal and manifesting as saturation effects. Signal waveforms distort during severe PMT saturation. Furthermore, when two light signals arrive within a short time interval, voltage deviations induced by the first signal persist, causing additional suppression of the subsequent signal.

\begin{figure}[htbp]
\centerline{\includegraphics[width=\linewidth]{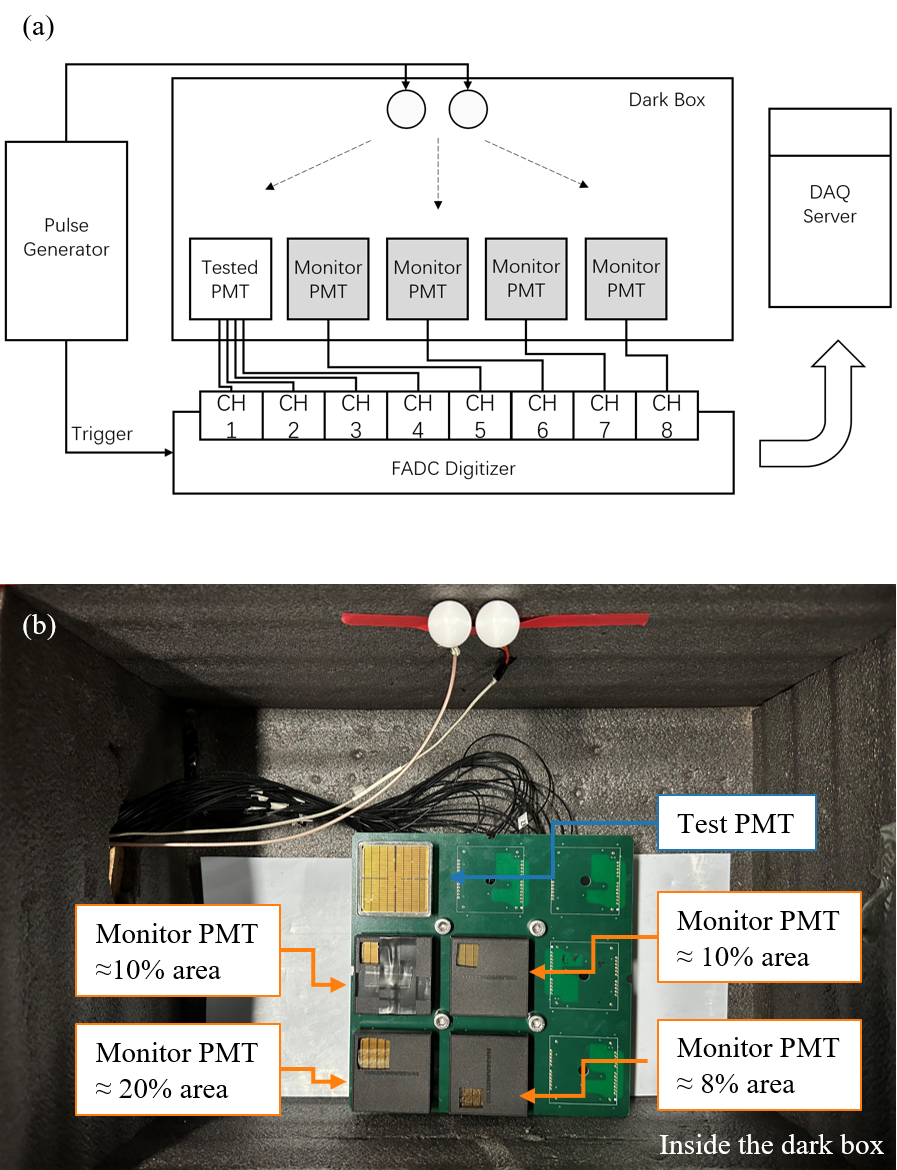}}
\caption{(a) Schematic of the bench-test setup. 
(b) Picture of the test setup. The base PCB accommodates a 3$\times$3 PMT array. 5 PMTs are placed inside, including 1 test PMT and 4 monitors.}
\label{fig:setup}
\end{figure}

Mitigation of saturation and suppression effects can be achieved by reducing PMT gain or optimizing the base circuit—specifically adding desaturation capacitors. 
However, given signal-to-noise ratio constraints for SPE detection, optimizing the number of desaturation capacitors in the base circuit is the only viable approach. 
For instance, capacitors C1–C4 are connected in parallel with dynodes DY8, DY9, DY10, and the anode, respectively, to maintain stable inter-stage voltages under large-signal conditions. While such capacitors effectively alleviate saturation, their configuration requires careful optimization to avoid excessive radioactivity.

Finally, three base circuits (BASE-1, BASE-2, and BASE-3) have been designed and implemented for bench testing, as detailed in Table~\ref{tab:base_config} and Fig.~\ref{fig:base_design_v2}. Consequently, subsequent testing focuses exclusively on desaturation capacitor placement.

\begin{table}[htbp]
\renewcommand{\arraystretch}{1.2} % 调整行间距
\caption{Three PMT Base Designs (Dynamic Range Discussed Later)}
\label{tab:base_config}
\centering
\setlength{\tabcolsep}{6pt}
\begin{tabular}{>{\centering\arraybackslash}p{42pt} >{\centering\arraybackslash}p{150pt}>{\centering\arraybackslash}p{42pt}}
\toprule
Base & Description & Dynamic range \\
\midrule
BASE-1 & Full desaturation (C1-C4) for max dynamic range & 286~nC \\
BASE-2 & Remove C1 to balance dynamic range and radioactivity & 83~nC\\
BASE-3 & Remove C1-C2 for minimal radioactivity & 25~nC\\

\bottomrule
\end{tabular}
\end{table}

\section{Performance test}
\label{test chapter}

\subsection{Bench test setup}

As shown in Fig.~\ref{fig:setup}, we set up a dedicated bench test for the PMT base to evaluate saturation and suppression effects.
The system consists of a dark box, five R12699 PMTs, a Tektronix AFG 31252 arbitrary waveform generator~\cite{TektronixAFG31252}, two blue LEDs embedded in PTFE spherical diffusers, a CAEN R8033DM high-voltage power supply~\cite{CAENR8033}, and a set of electronics together with a data acquisition (DAQ) system~\cite{He2021Digitizer}.

The five PMTs are arranged as follows: one functions as the test PMT, equipped with the base under study.
The saturation response is evaluated using the sum of the charges from all four channels. 

In order to monitor the light intensity over a wide dynamic range, four monitor PMTs are installed next to the tested PMT. These monitor PMTs employ BASE-1 and are equipped with black masks covering the photocathode, reducing their effective light-sensitive area to 8–20\% of full size, so that for any given LED signal, at least one of them operates in its linear regime.
In the analysis, the monitor PMT with the highest signal-to-noise ratio that still lies in the linear regime is used to offer a reference.

The true response of the tested PMT is reconstructed by multiplying the charge of the chosen monitor PMT by the ratio of the tested PMT response to the monitor response measured in the linear regime.

The Tektronix AFG31252 drives the LEDs. Its two phase-synchronized channels generate timing correlated light-outputs, producing either S2-like waveforms using a 10~{\textmu}s-wide square waveform. 
The diffuser attached to each LED makes the light output as isotropic as possible. Light intensity is adjusted via the generator's output amplitude. The DAQ operates in external-trigger mode, synchronized with the generator, and digitizes the PMT signals at a sampling rate of 500~MHz~\cite{He2021Digitizer}. 
The DAQ digitizes the PMT waveforms with a dynamic range of $\pm1.08$~V with 14-bit resolution, corresponding to a voltage resolution of 0.132~mV.

In this work, a total of eight R12699 PMTs are included in the test, labeled A through H. 
PMT-A operating at a gain of $2.7\times10^6$ serves as the main target device.
SPE charge calibration precedes and follows each bench test to monitor PMT gain stability. Gain remained stable throughout all measurements. The collected charge is analyzed to determine the PMT gain, which is then used to infer input light intensity.

\subsection{Saturation}
\begin{figure}[htbp]
\centerline{\includegraphics[width=\linewidth]{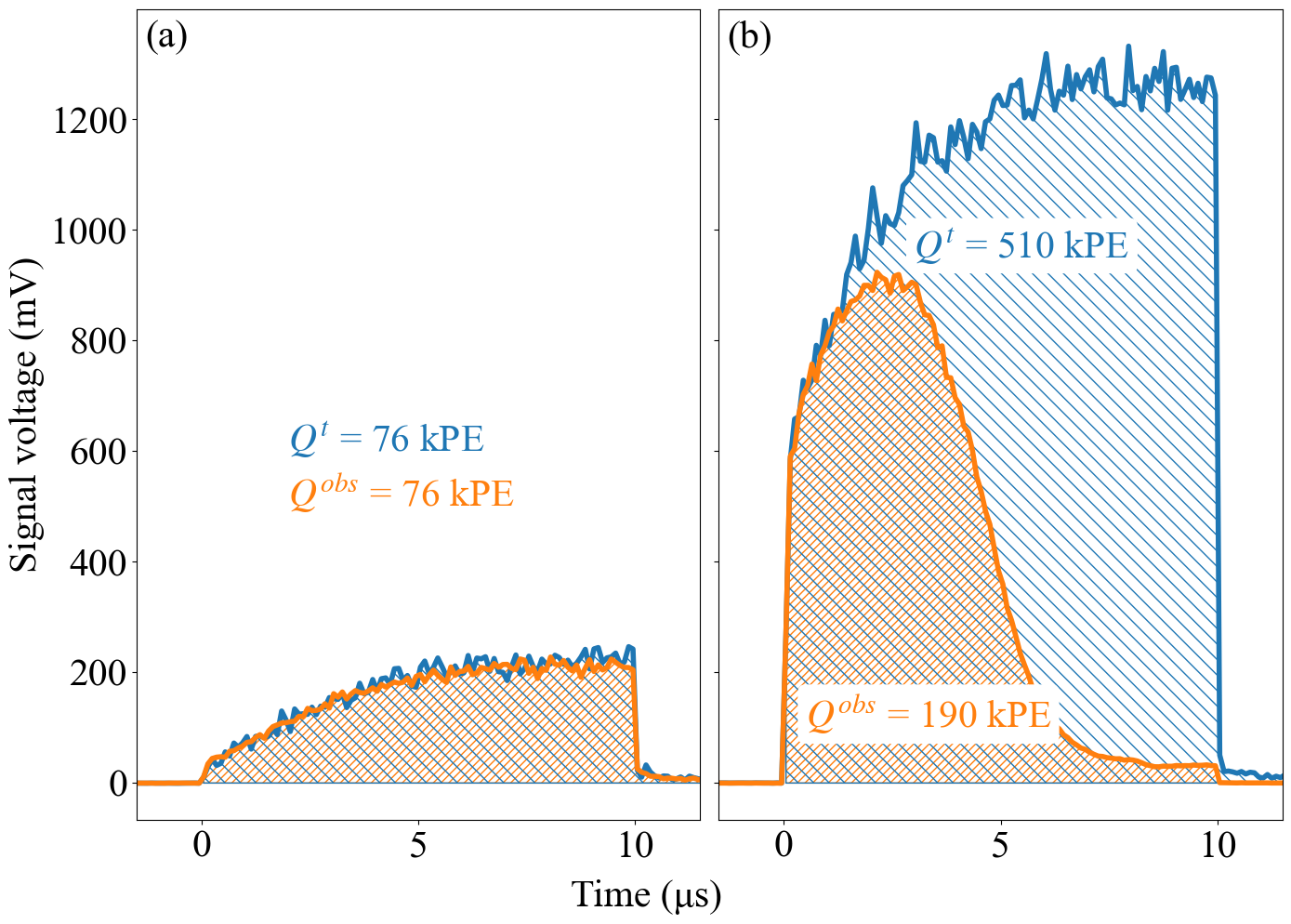}}
\caption{PMT waveforms from bench test. The orange and blue waveforms represent the recorded signal from the test PMT and the reconstructed input signal from the monitor PMT, respectively. (a) the observed charge $Q^{obs}$ matches the true input charge $Q^{t}$, (b) shows Signal with severe saturation effect.}
\label{fig:sat_wave}
\end{figure}

Saturation effect manifests as waveform distortion and reduced PMT charge response. In Fig.~\ref{fig:sat_wave} (a), the observed charge $Q^{obs}$ matches the true input charge $Q^{t}$, which is reconstructed from the monitor PMT.
In Fig.~\ref{fig:sat_wave}(b), with high light input, severe saturation occurs, waveforms distort significantly, and $Q^{obs}$ is reduced by 63\% compared to $Q^{t}$.

\begin{figure}[htbp]
\centerline{\includegraphics[width=\linewidth]{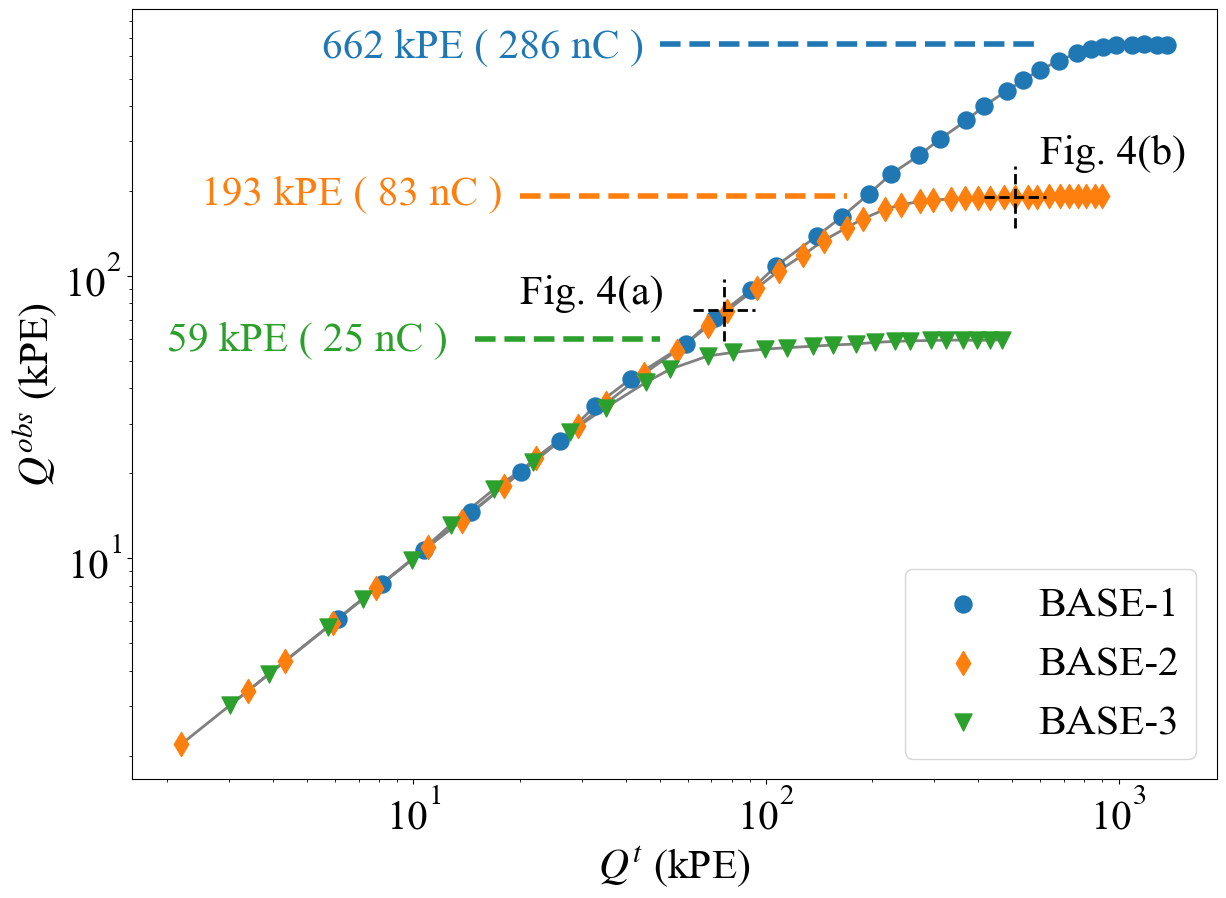}}
\caption{The saturation response curves of PMT-A with different base circuits. The black markers denote the two specific cases presented in Fig.~\ref{fig:sat_wave}.}
\label{fig:sat_curve}
\end{figure}

Fig.~\ref{fig:sat_curve} presents the saturation response curves of $Q^{obs}$ as a function of $Q^t$ for three bases.
BASE-1 exhibits the widest dynamic range, with a maximum $Q^{limit}$ of 662~kPE (286~nC).
Compared to BASE-1, $Q^{limit}$ of BASE-2 drops to one-third (83~nC), while that of BASE-3 decreases to one-tenth (25~nC).
This limit arises from the complete depletion of the charge stored in C$_\mathrm{eq}$. 
Notably, the $Q^{limit}$ ratio between BASEs directly indicates that the amplification factor of the subsequent dynodes is 3.4.
Following this ratio, the charge at Dy9--D10 and DY8--DY9 reach about 24~nC and 7~nC at $Q^{limit}$ for BASE-2, respectively.
This value correlates strongly with the charge stored in C$_\mathrm{eq}$ (8~nC), confirming C$_\mathrm{eq}$’s role in limiting dynamic range.

\begin{figure}[htbp]
\centerline{\includegraphics[width=\linewidth]{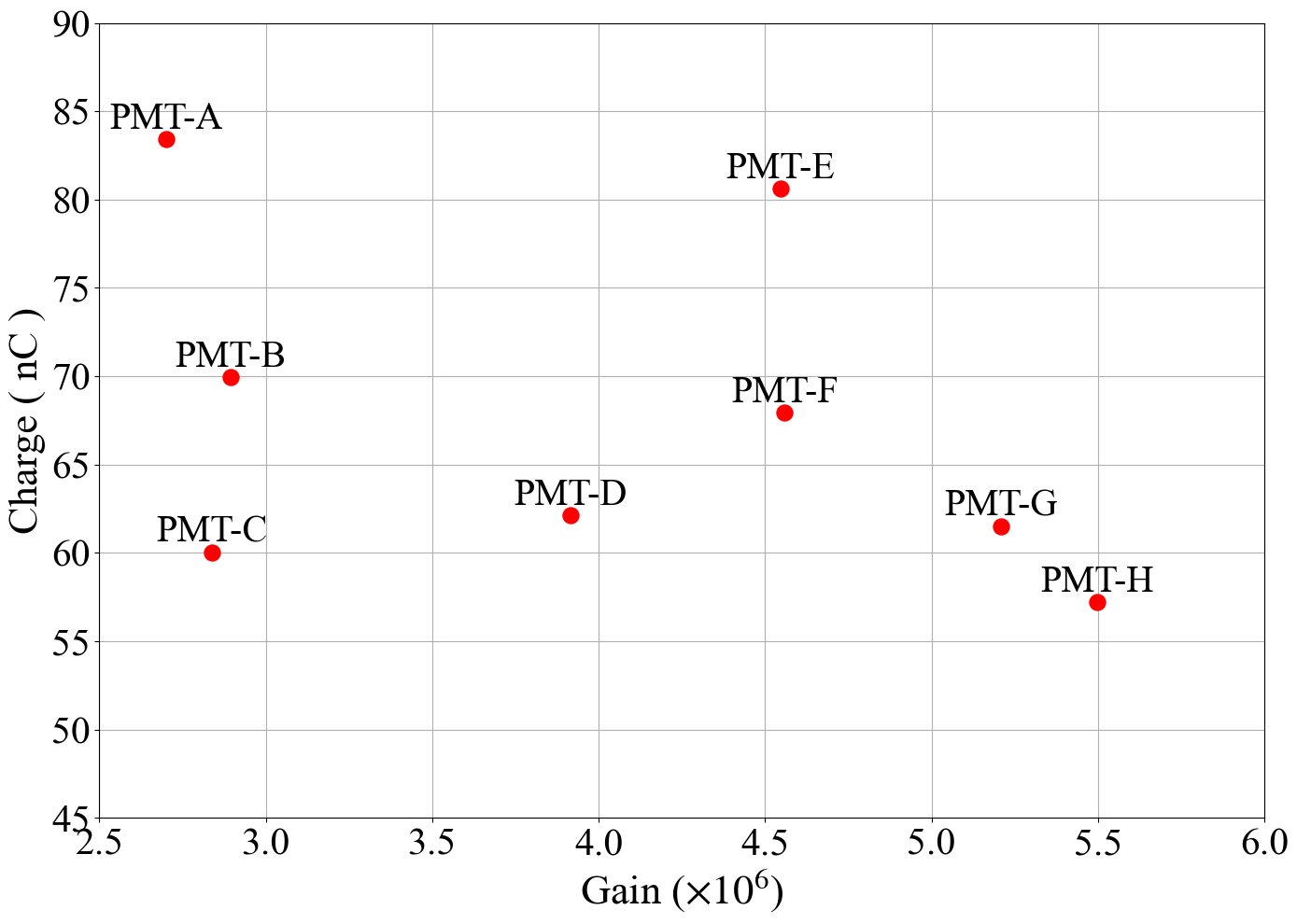}}
\caption{Saturation limits for multiple PMTs with BASE-2 at different gains spanning 2.7 to $5.5\times10^6$.}
\label{fig:gain_limit}
\end{figure}

As shown in Fig.~\ref{fig:gain_limit}, PMTs with BASE-2 deliver a measured $Q^{limit}$, in most cases, above 60~nC, thereby exceeding twice the aforementioned dynamic-range requirement. In the remainder of this work, we use BASE-2 as the default base configuration unless otherwise specified.

Moreover, established correction methods can further desaturate PMT signals by utilizing waveform information~\cite{Luo2024}.

\subsection{Suppression}

\begin{figure}[htbp]
\centerline{\includegraphics[width=\linewidth]{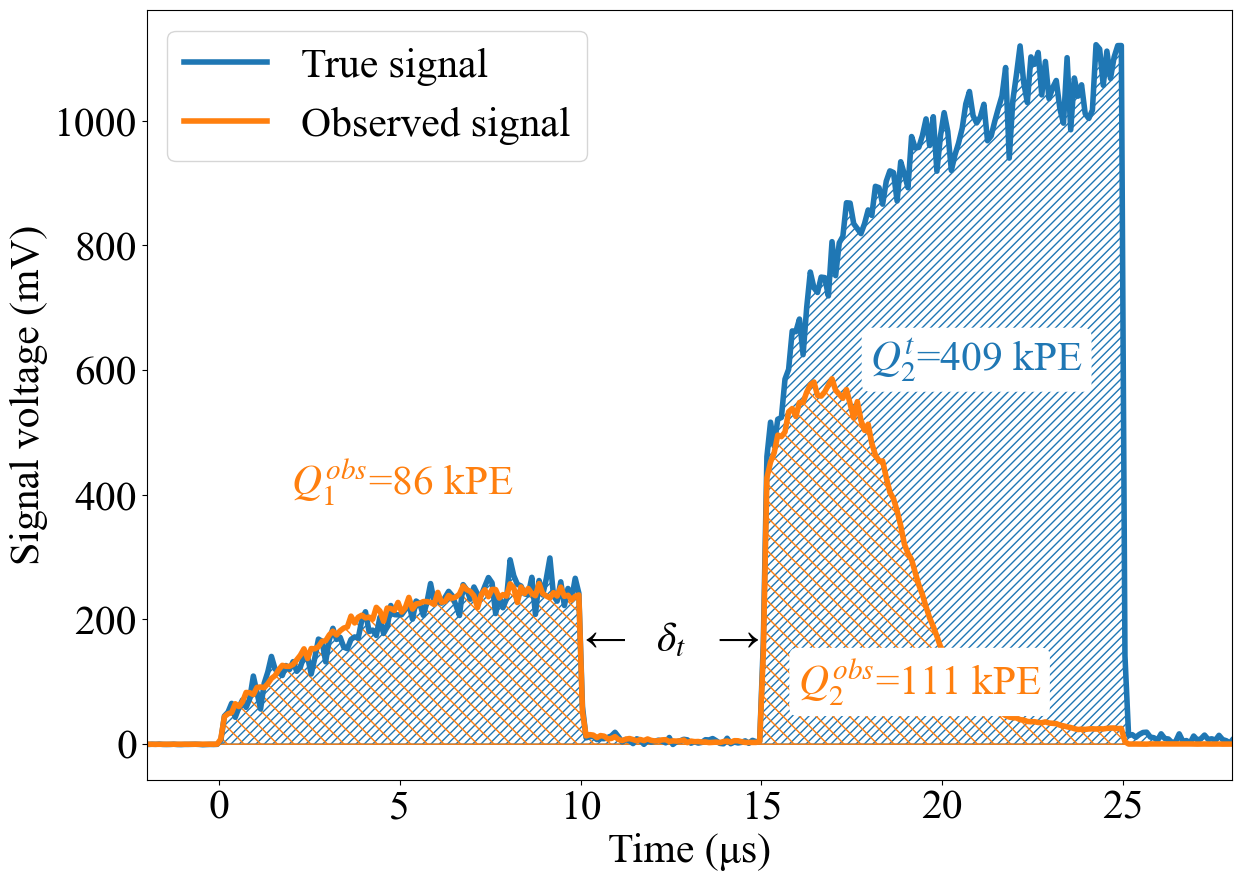}}
\caption{Suppression response waveforms of PMT-A with BASE-2.
The orange lines represent the recorded signal from the test PMT, and the blue lines represent the reconstructed true input signal from the monitor PMT. $\delta_t$ here is 5~{\textmu}s.}
\label{fig:sup_wave}
\end{figure}

We characterize the suppression effect using the same bench-test setup with PMT-A and BASE-2.
Fig.~\ref{fig:sup_wave} illustrates a suppression event where two 10-{\textmu}s square pulses are separated by a time interval $\delta_t$ of 5~{\textmu}s.
In the first waveform, the observed signal corresponds well with the true signal, while the second waveform shows clear distortion, and its observed charge $Q_{2}^{\mathrm{obs}}$ is only 111~kPE, showing a significant reduction compared with the true input charge $Q_{2}^{\mathrm{t}}$ of 409~kPE.

\begin{figure}[htbp]
\centerline{
\includegraphics[width=\linewidth]{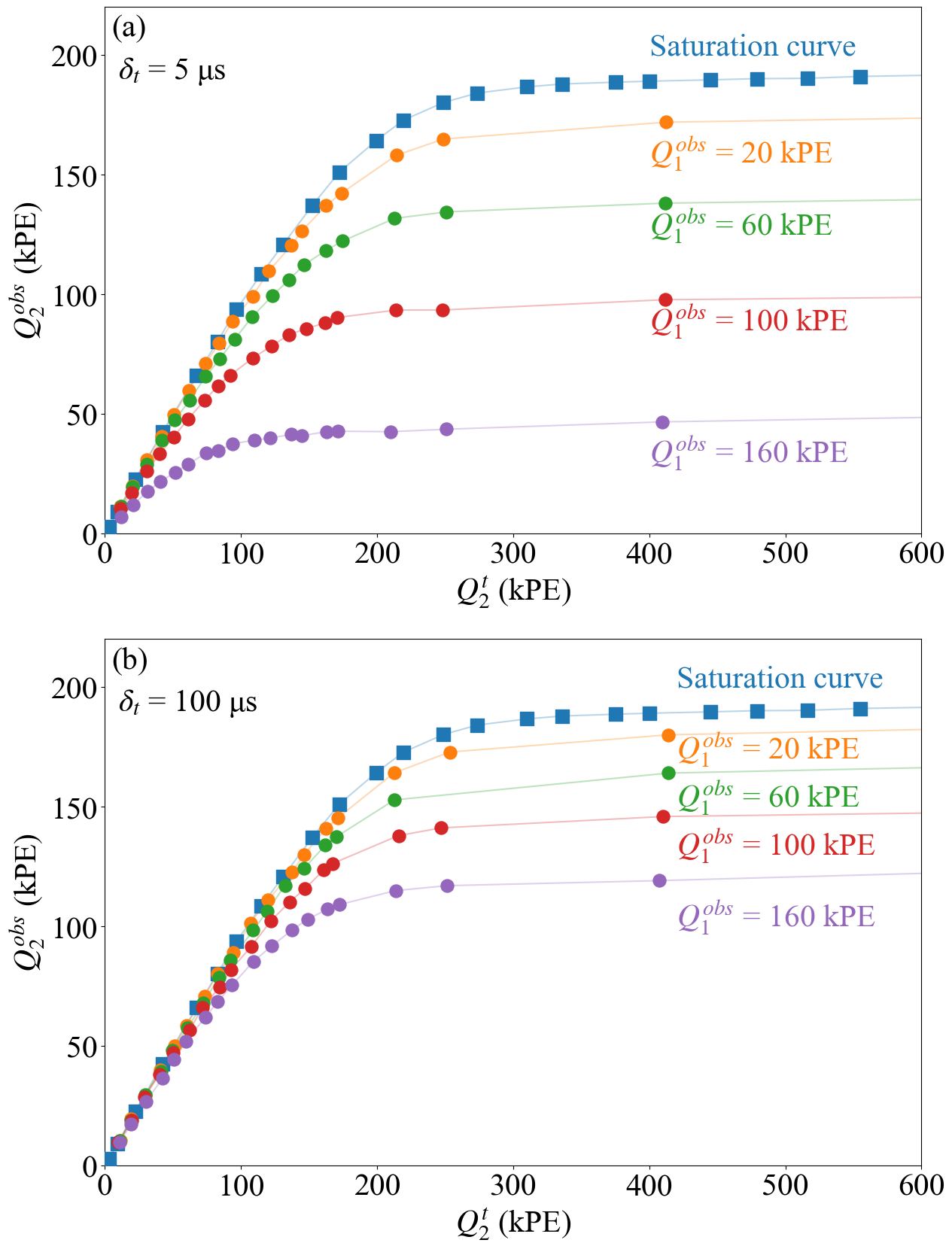}

}
%\centerline{}
\caption{PMT suppression curves of different $Q_1^{obs}$ for $\delta_t$ of 5~{\textmu}s (a) and 100~{\textmu}s (b), the saturation curve is overlaid.}
\label{5vs100}
\end{figure}

The suppression effect depends on both $Q_{1}^{\mathrm{obs}}$ and $\delta_t$. As shown in Fig.~\ref{5vs100}, for a given $\delta_t$, the $Q_2^{\mathrm{obs}}$ response follows a saturation-like curve, approaching a limiting value that is inversely proportional to $Q_1^{\mathrm{obs}}$, reflecting the capacitor discharging induced by $Q_1^{\mathrm{obs}}$.
Conversely, for a given $Q_1^{\mathrm{obs}}$, a longer $\delta_t$ mitigates the suppression effect. By comparing Fig.~\ref{5vs100}~(a) and (b), at $\delta_t = 100$~{\textmu}s, the $Q_2^{\mathrm{obs}}$ response curve corresponding to a given $Q_1^{\mathrm{obs}}$ exhibits a higher limiting value and approaches the saturation curve more closely, indicating enhanced capacitor recharging.

\begin{figure}[htbp]
\centerline{\includegraphics[width=\linewidth]{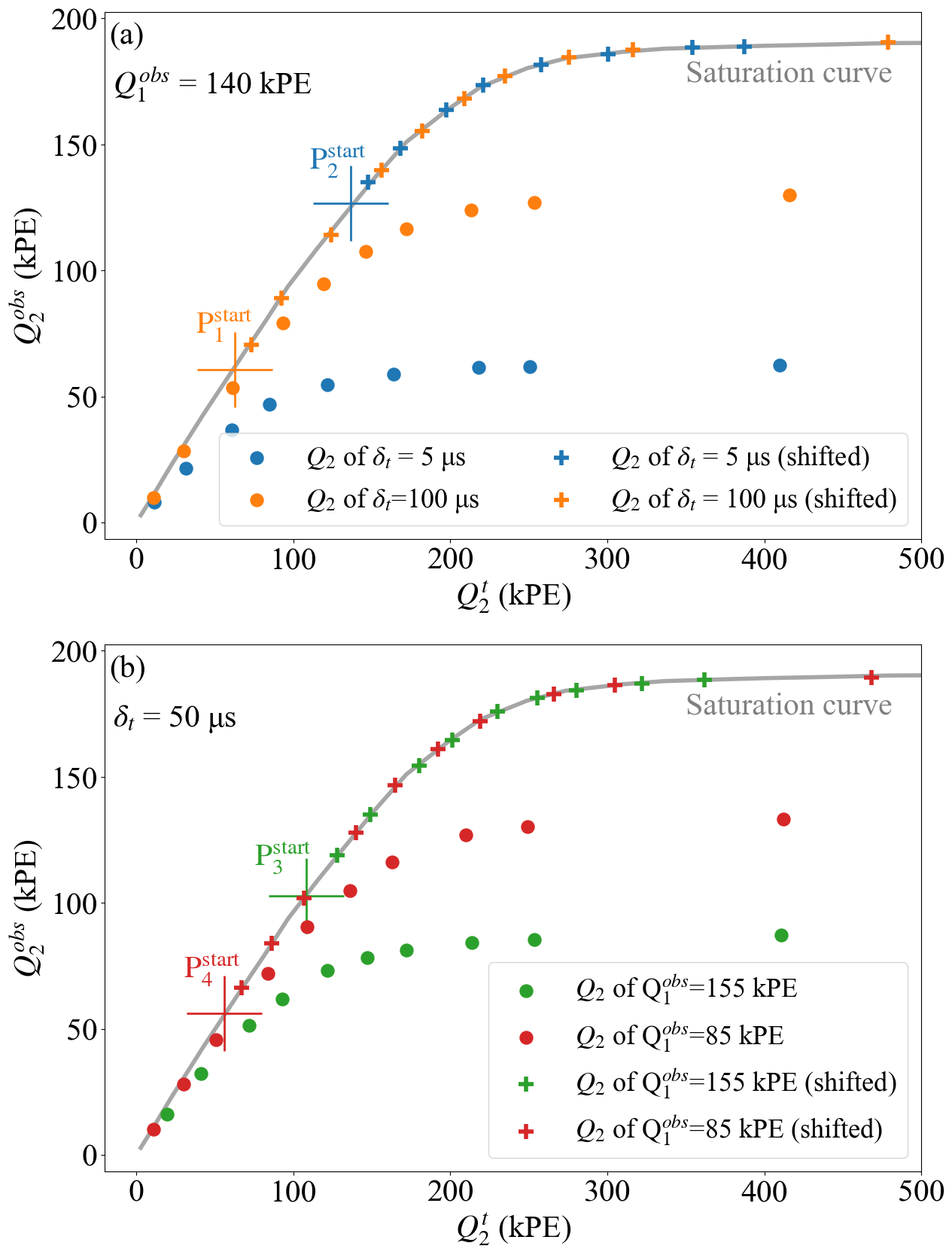}}
\caption{Comparison between saturation and suppression response. (a) compares response curves at given $Q_1^{obs}$ of 140~kPE for two different $\delta_t$ of 5~{\textmu}s and 100~{\textmu}s. (b) displays curves at given $\delta_t$ of 50~{\textmu}s for two different $Q_1^{obs}$ of 83~kPE and 155~kPE. The definition of ${\mathrm{P}}^{\mathrm{start}}$ can be found in the text.}
\label{fig:correlation}
\end{figure}

Compared to the saturation, the first waveform shifts the subsequent response, thereby altering the onset of the charge response in the saturation curve.
To clarify the relationship between suppression and saturation, we propose a novel shift mode based on the start point.
To determine the start point: firstly, compare the difference between the maximum of $Q_2^{obs}$ and $Q^{limit}$; subsequently, intersect this difference with the saturation curve, and the intersection point thus obtained is defined as the start point.
Suppression-induced response curves can be aligned with the saturation curve by introducing appropriate response shifts at the aforementioned start points
(e.g. ${\mathrm{P}}_{1}^{\mathrm{start}}$ and ${\mathrm{P}}_{2}^{\mathrm{start}}$, or ${\mathrm{P}}_{3}^{\mathrm{start}}$ and ${\mathrm{P}}_{4}^{\mathrm{start}}$). 
Fig.~\ref{fig:correlation}~(a) compares response curves at given $Q_1^{obs}$ of 140~kPE for two different $\delta_t$ of 5~{\textmu}s and 100~{\textmu}s. 
For larger $\delta_t$, the start point shifts toward zero; as $\delta_t \to \infty$, the start point shifts to zero, thereby restoring the saturation scenario.
Fig.~\ref{fig:correlation}~(b) displays curves at given $\delta_t$ of 50~{\textmu}s for two different $Q_1^{obs}$ of 83~kPE and 155~kPE. 
When $Q_1^{obs}=0$, the start point also shifts to zero.
Therefore, the physical mechanism of the suppression effect resembles saturation.

\begin{figure}[htbp]
\centerline{\includegraphics[width=\linewidth]{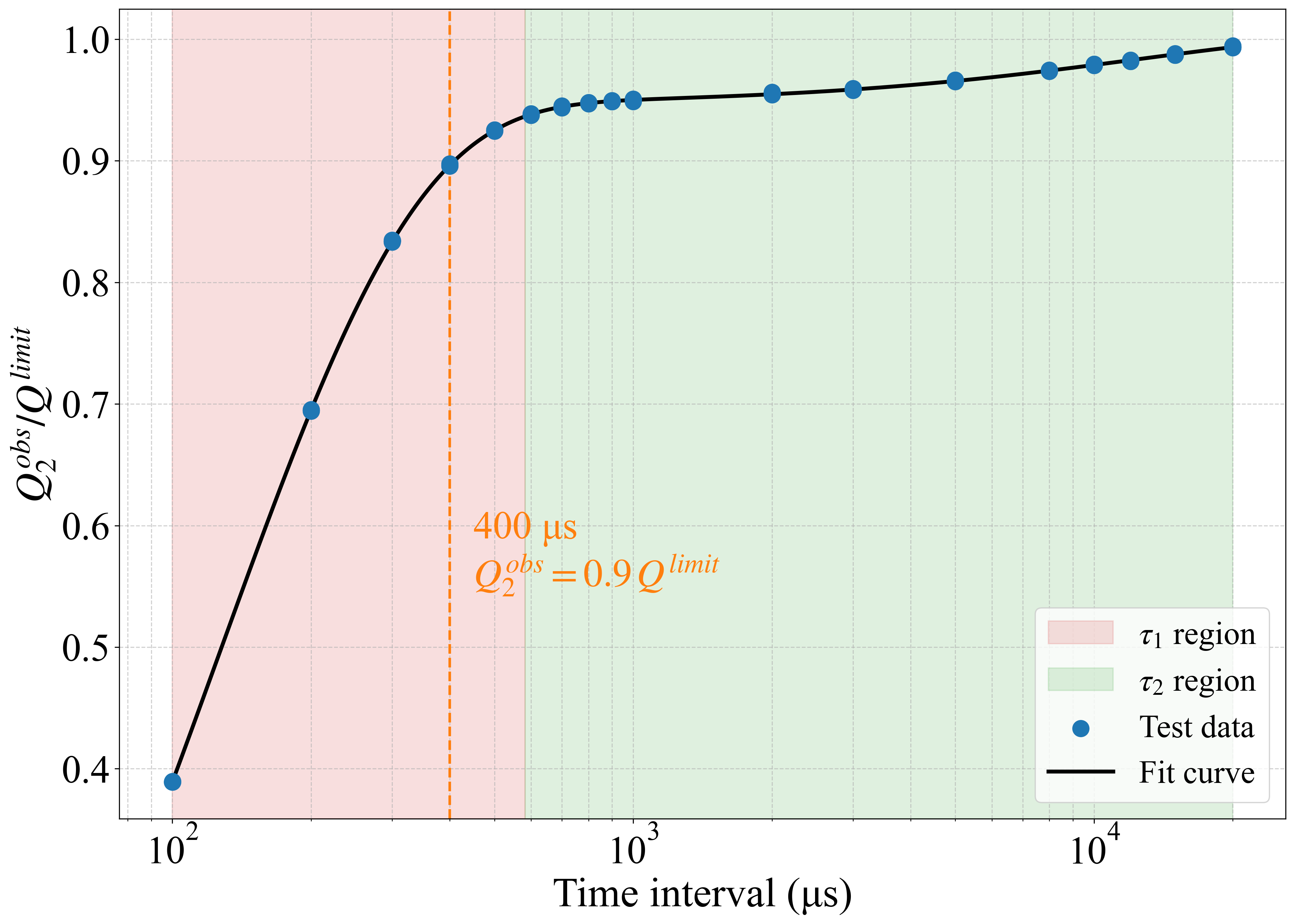}}
\caption{Recovery value of the second charge with $Q_1^{obs} = Q^{limit}$. The recovery time consists of two components (a fast component shown as the red region and a slow component shown as the green region). The fitted function is $y=(0.53-0.69\cdot\mathrm{e^{-x/125.85}})\times(1.9-0.11\cdot\mathrm{e^{-x/12365.42}})$.}
\label{fig:totalrecovery}
\end{figure}

For a given $Q_1^{obs} = Q^{limit}$, we further investigated the recovery time of the suppression effect.
Fig.~\ref{fig:totalrecovery} illustrates the relationship between $\frac{Q_2^{obs}}{Q^{limit}} $ and $\delta_t$: the recovery time consists of two components (fast shown as the red region and slow shown as the green region), corresponding to the equivalent capacitance C$_\mathrm{eq}$ and desaturation capacitors, respectively.
The fitting function is defined as the product of two capacitor charging functions, each associated with the respective RC time constant of the two capacitors.
%(i.e., $\tau_\mathrm{eq}$ for C$_\mathrm{eq}$ and $\tau_{\mathrm{C}}$ for desaturation capacitors).
Since the response of the PMT is non-linear with respect to the capacitor voltage, the fitting result should only be taken as a reference.
After approximately four relaxation times (fitted $\tau_{1}$=125~$\mu$s, C$_\mathrm{eq}$ dominated), $Q_2^{obs}$ reaches 90\% of $ Q^{limit}$; 
full recovery of the suppression effect requires 20~ms, governed by the slow relaxation of desaturation capacitors (fitted $\tau_2$ = 12~ms, desaturation capacitor dominated).
Consequently, for LXe detectors, the 20~ms period following a large signal should be treated as the dead time, as the PMT has not yet fully recovered during this interval.
Additionally, the parameters derived from the fitted curve indicate that the characteristic capacitance of C$_\mathrm{eq}$ is approximately \( 0.1\,\text{nF} \), which is consistent with the value measured by the LCR meter.

\begin{figure}[htbp]
\centerline{\includegraphics[width=\linewidth]{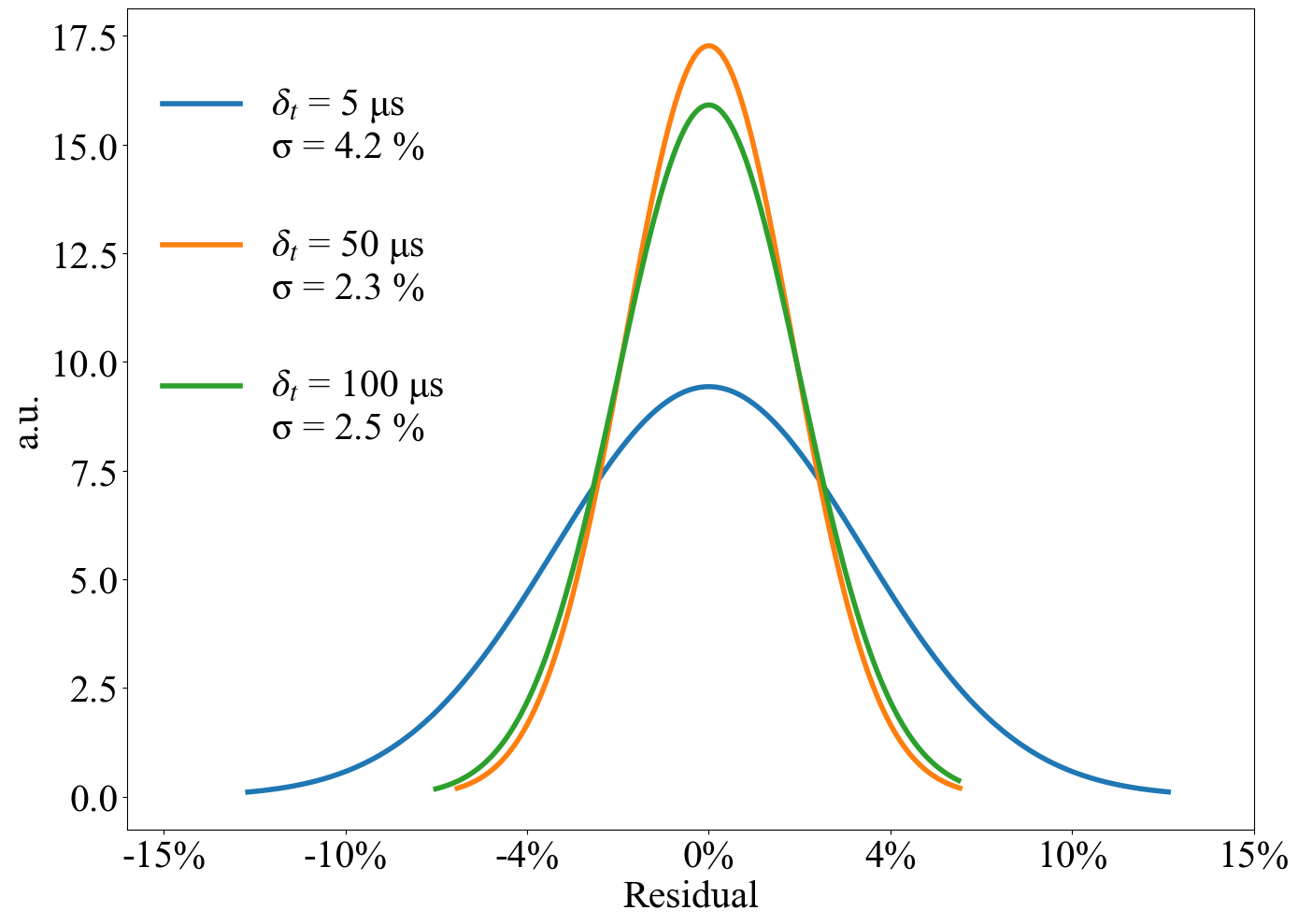}}
\caption{Distributions of relative residual after correction of the second charge for $\delta_t$ of 5~{\textmu}s, 50~{\textmu}s, and 100~{\textmu}s, where the residual is quantified by $\mathcal{R} = \frac{Q_2^{c} - Q_2^{t}}{Q_2^{t}}$.}
\label{fig:error_gauss}
\end{figure}

Given values of $Q_1^{obs}$ and $\delta_t$, the response starting point can be determined, enabling correction for the suppression effect by following the saturation response curve. A correction factor $\mathcal{F}$ is defined as the ratio of the charge on the saturation curve to the observed charge of the second waveform. Fig.~\ref{fig:error_gauss} shows three cases of $\delta_t$ of 5~{\textmu}s, 50~{\textmu}s, and 100~{\textmu}s. The correction performance is quantified by the relative residual $\mathcal{R} = \frac{Q_2^{c} - Q_2^{t}}{Q_2^{t}}$, where $Q_2^{t}$ denotes the true charge of the second signal and $Q_2^{c}$ is the corrected charge ($Q_2^{c} = \mathcal{F} \times Q_2^{obs}$). For shorter $\delta_t$, suppression intensifies, leading to less reliable corrections. These results support developing a general PMT correction database, offering an additional tool to recover suppressed signals in future experiments.

\section{Base circuit simulation}
\label{simu chapter}

\begin{figure*}[htbp]
\centerline{\includegraphics[width=\textwidth]{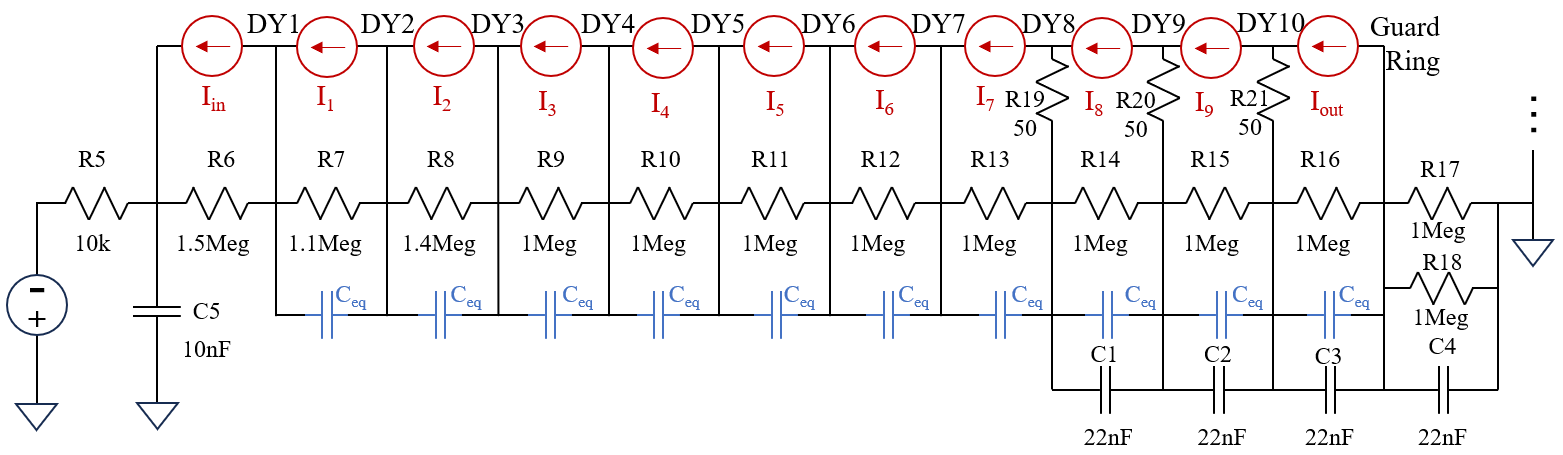}}
\caption{LTSpice circuit model of PMT-A with BASE-2, including the dynode chain, current sources, and measured inter-dynode capacitances.}
\label{fig:spice_model}
\end{figure*}

Circuit simulations provide deeper insight into the mechanisms of saturation and suppression effects. We employ the LTSpice software~\cite{ltspice24120} to simulate the base circuit connected with R12699. The model of simulation is shown in Fig.~\ref{fig:spice_model}, where $\mathrm{I_{in}}$ represents the current emitted from the PMT photocathode. Current sources are placed between adjacent dynodes to model signal transmission behavior, including voltage division, electron multiplication, and electron transport.
The output signal $\mathrm{I}_\mathrm{out}$ corresponds to the current extracted from the final dynode. The simulation also includes the equivalent capacitance $\mathrm{C}_\mathrm{eq}$ of 0.1~nF between adjacent dynodes.

A data-driven method is employed to determine the current sources from $\mathrm{I}_\mathrm{1}$ to $\mathrm{I}_\mathrm{out}$ based on our bench test.
Ideally, electron multiplication factors between dynodes are denoted by $\delta_{n}$ (from $\delta_{1}$ to $\delta_{10}$), with specific values provided by Hamamatsu Corporation as reference.
At $-$1000~V voltage, the gain of PMT-A $G = \prod_{n=1}^{10}\delta_{n}$ is $2.7 \times 10^{6}$.

In practice, charge transport induces a proportional reverse bias $\Delta V_n$ at each dynode stage, reducing the effective multiplication voltage, lowering the gain by $\Delta \delta_n$, and introducing nonlinearity, particularly in the later stages.

Thus, the output signal can be expressed as:

\begin{equation}
\mathrm{I}_\mathrm{out} = \mathrm{I}_\mathrm{in} \times \prod_{n=1}^{10} (\delta_n - \Delta \delta_n),
\end{equation}

where $\Delta \delta_n$ is modeled with $A \times \Delta V_n ^B$ in the simulation. The proportional coefficient $A$ and the constant $B$ are tuned through bench test data.

\begin{figure}
    \centering
    \includegraphics[width=1\linewidth]{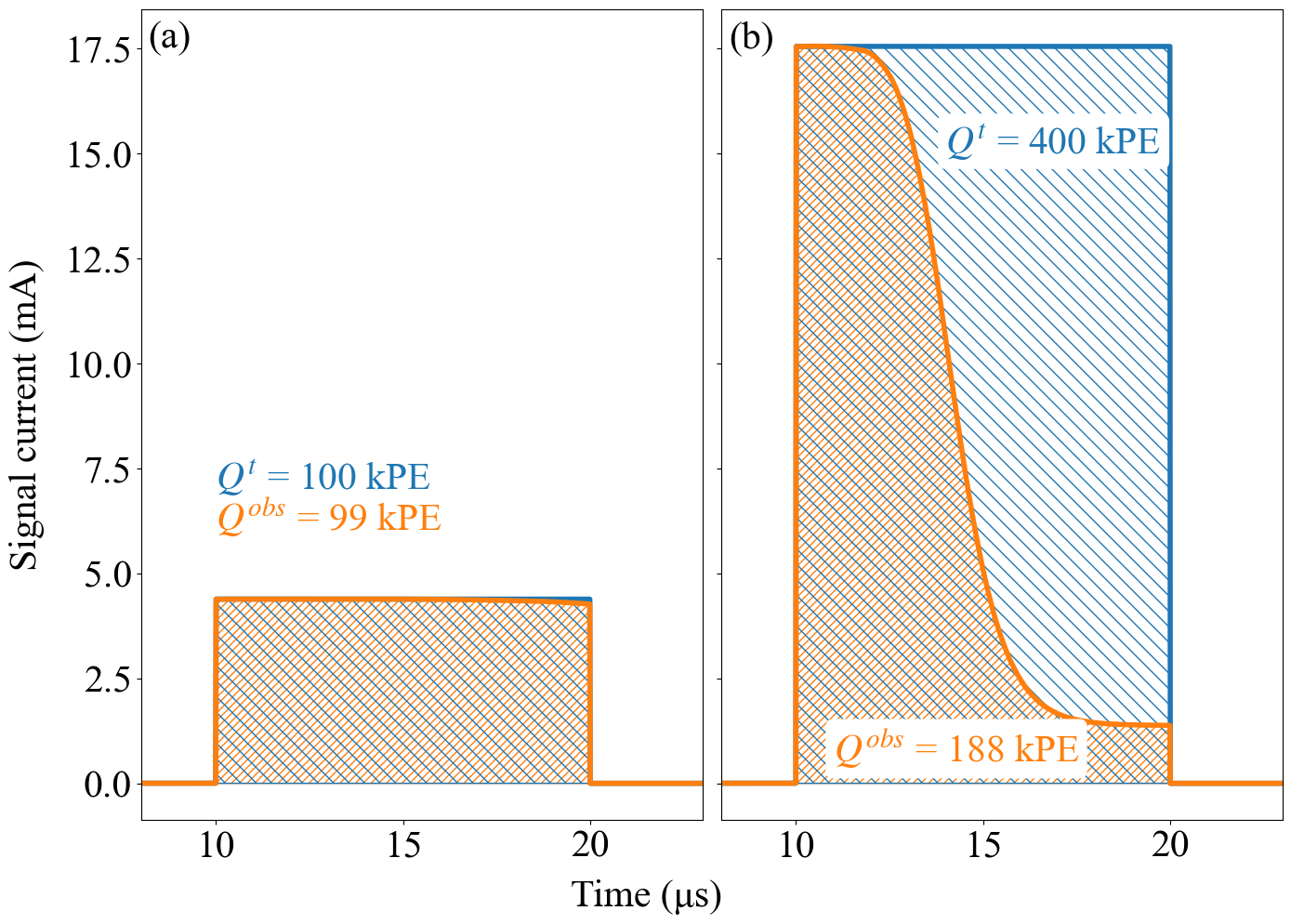}
    \caption{Illustration of the simulated waveforms. 
(a) Simulation without saturation. 
(b) Simulation with saturation.}
    \label{fig:sat_simu}
\end{figure}

\begin{figure}
    \centering
    \includegraphics[width=1\linewidth]{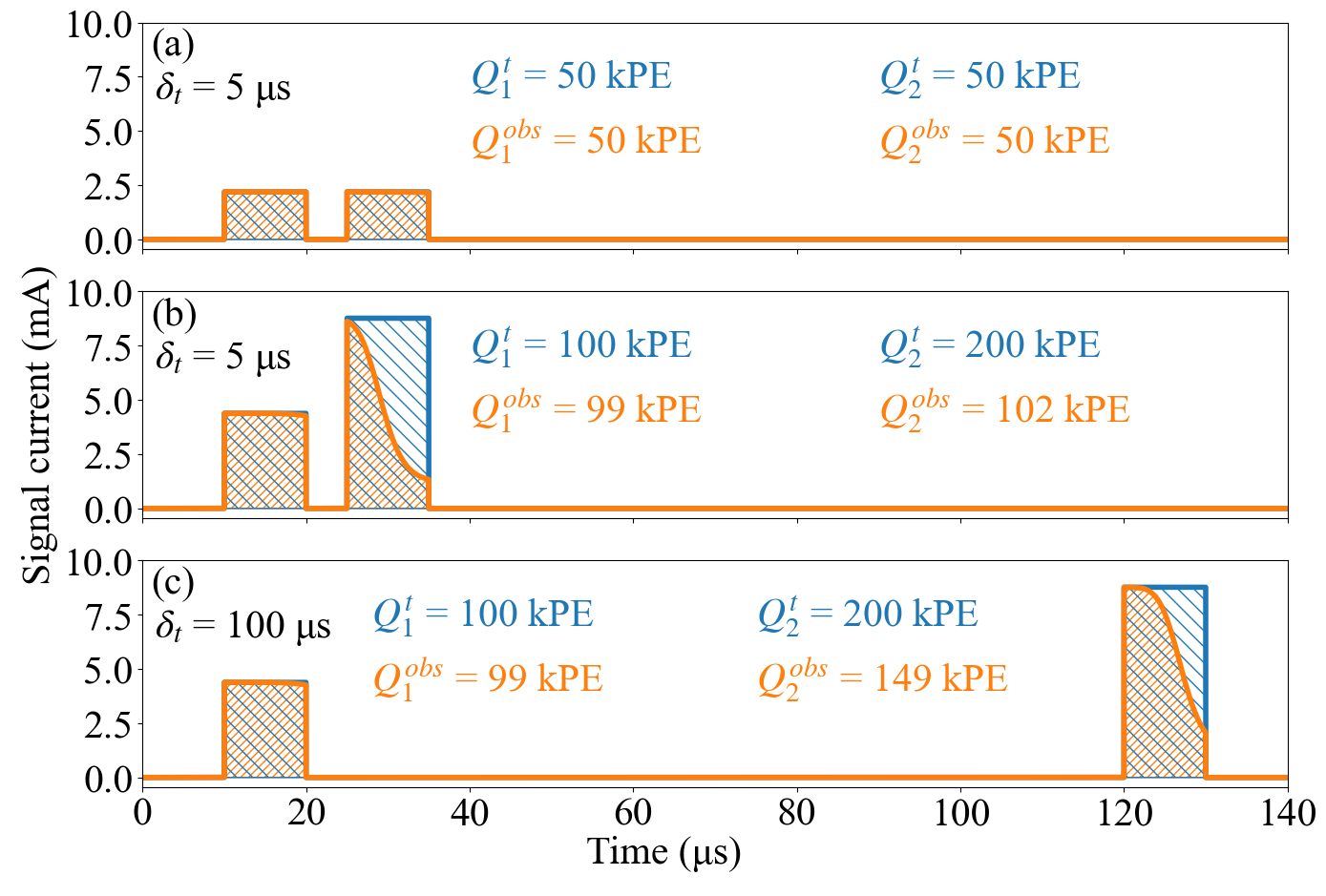}
    \caption{Simulated examples for suppression for two consecutive inputs. 
(a) For small signals ($Q_1^{t}$ = 50~kPE, $Q_2^{t}$ = 50~kPE) with $\delta_{t}$ = 5~{\textmu}s. 
(b) For larger signals ($Q_1^{t}$ = 100~kPE, $Q_2^{t}$ = 200~kPE) with the same interval. 
(c) With a longer interval of $\delta_{t}$ = 100~{\textmu}s.}
    \label{fig:sup_simu}
\end{figure}

The simulation reproduces the PMT response, including characteristic saturation and suppression effects.
Fig.~\ref{fig:sat_simu} shows two examples of simulated PMT waveforms; no waveform distortion occurred for an input signal of 100~kPE with a width of 10~{\textmu}s. However, under a larger input of 400~kPE, significant saturation distortion is observed, yielding an observed charge $Q^{\mathrm{obs}}$ of 188~kPE.
Fig.~\ref{fig:sup_simu} shows the characteristic waveforms of the suppression effect, two input square pulses with a width of 10~{\textmu}s are applied.
Fig.~\ref{fig:sup_simu} (a) shows an example under the small signal situation, where no suppression effect is observed.
Fig.~\ref{fig:sup_simu} (b) and (c) demonstrate the examples of suppression response with different time intervals $\delta_t$ between the two signals. In the case of $\delta_t$~=~5~{\textmu}s, a significant suppression effect is observed.
By increasing $\delta_t$ to 100~{\textmu}s, the suppression effect is reduced, which is consistent with the bench test results in the experiment.

\begin{figure}[htbp]
\centerline{\includegraphics[width=\linewidth]{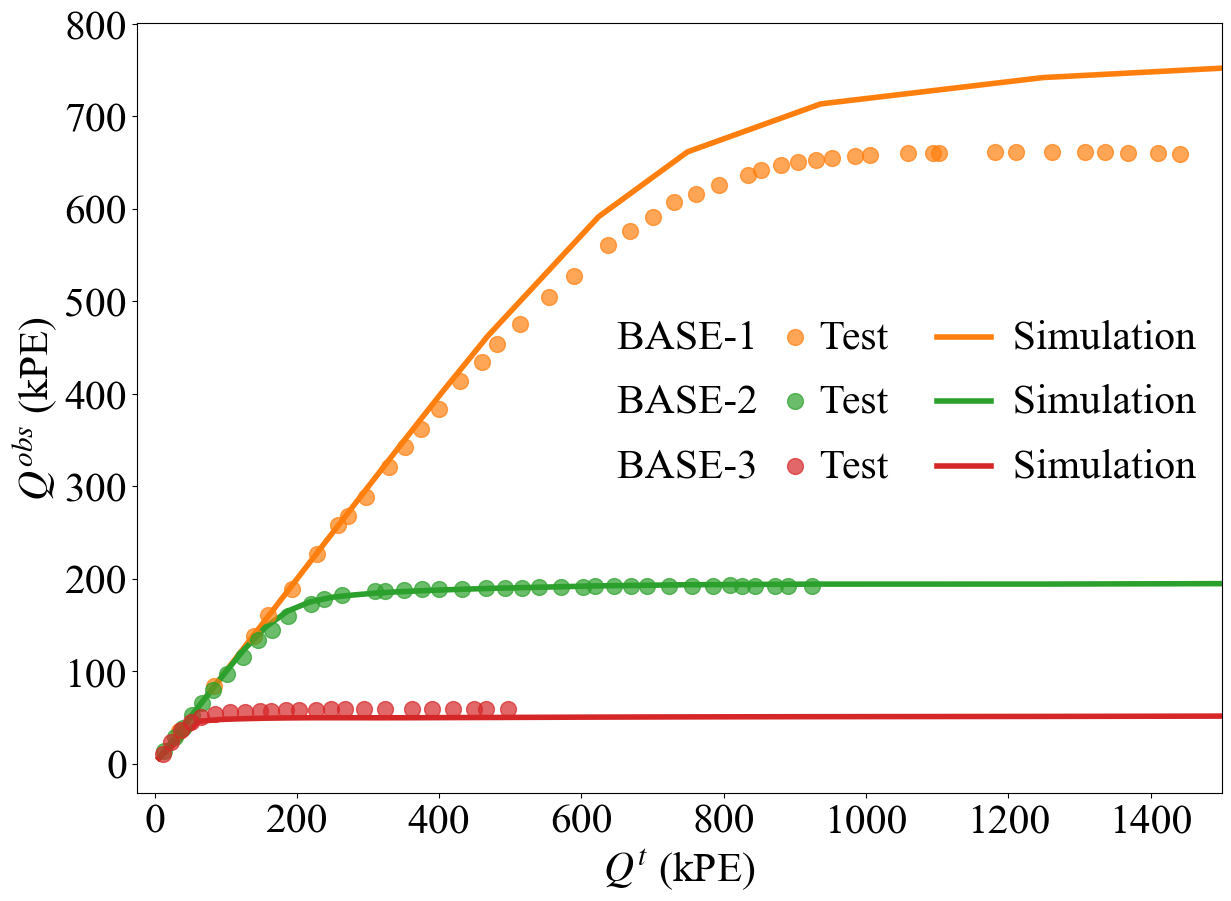}}
\caption{Comparison of saturation response curves between simulation and bench test results for PMT-A on BASE-1, BASE-2, and BASE-3.}
\label{fig:fig_a_saturation_only}
\end{figure}

\begin{figure}[htbp]
\centerline{\includegraphics[width=\linewidth]{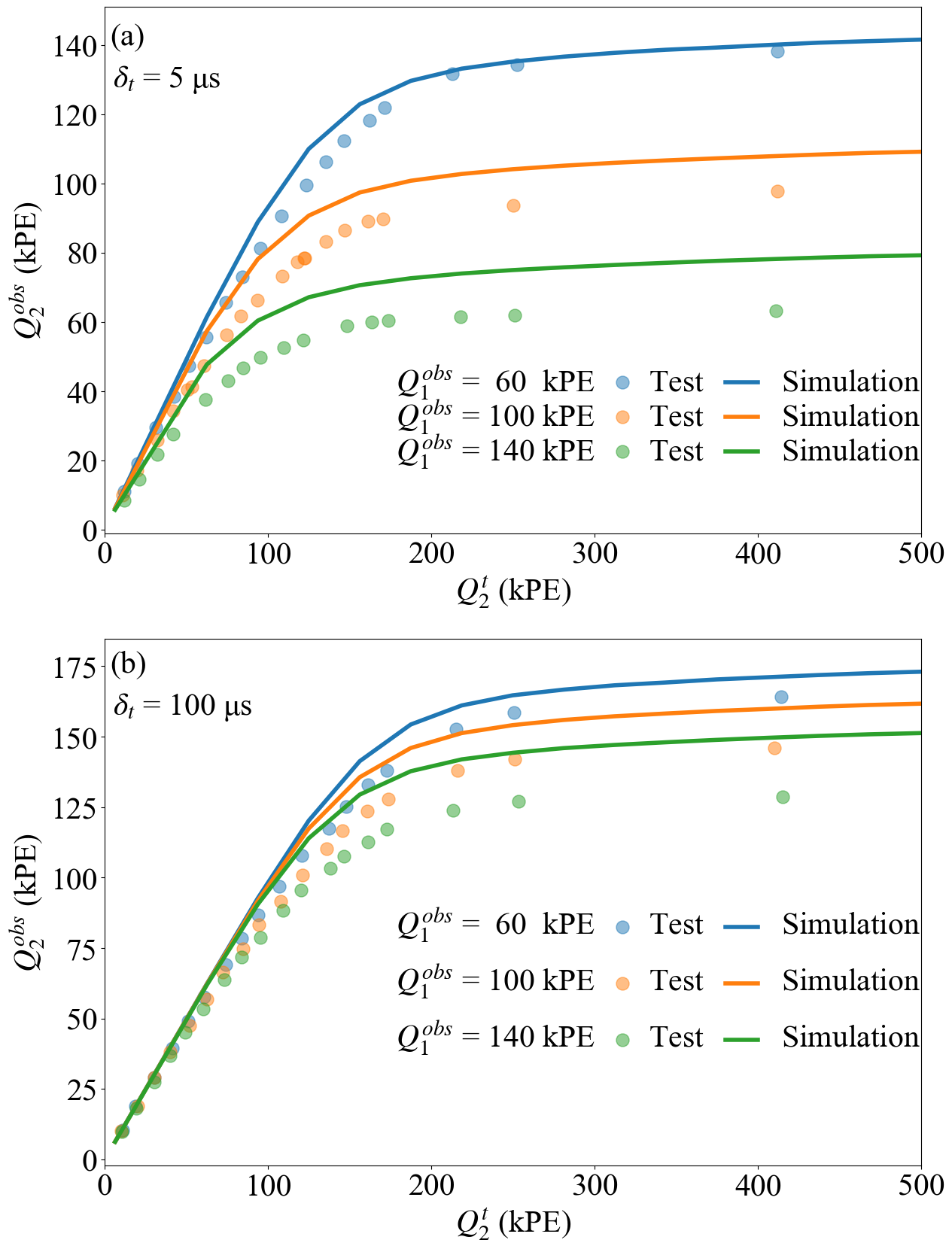}}
\caption{Comparison of suppression response curves between simulation and bench test results for PMT-A 
on BASE-2 for time intervals of 5~{\textmu}s and 100~{\textmu}s.}
\label{fig:fig_bc_pre_sub_only}
\end{figure}

Fig.~\ref{fig:fig_a_saturation_only} presents the saturation response curves for BASE-1, BASE-2, and BASE-3.
Simulation results closely match bench-test data in both waveform shapes and saturation limits, achieved solely through tuning factor $A$ using the BASE-2 configuration. 
Simulations closely match experimental behavior, with slight deviation only in the saturation limit for BASE-1.
It is also adopted in simulating the suppression effect.
Fig.~\ref{fig:fig_bc_pre_sub_only} shows the results for two different time intervals $\delta_t$. 
These discrepancies may originate from multiple sources, including response variations between PMT stages or finite response times of base circuit components. They can be mitigated through additional parameter tuning using experimental data. For example, we can tune the $\delta_n$ or replace the constants in $A \times \Delta V_n ^B$ by $A_n$ and $B_n$ to optimize the response simulation in each stage.
Finally, the circuit simulation successfully captures the essential physical mechanisms underlying the response, while also providing a platform for the design and performance evaluation of base circuits.

\section{Conclusion}
\label{sum chapter}

We report the design and performance study of the Hamamatsu R12699 PMT voltage divider base. The large-scale LXe experiments have demonstrated the capability to search for both WIMPs and NLDBD in the energy range from sub-keV up to several MeV.
The Hamamatsu R12699 PMT has the advantages of low background and high granularity, making it the preferred candidate for the next generation of detectors.
The optimized PMT base is a critical component of the detector for achieving a wide dynamic range.

Three bases with different numbers of desaturation capacitors were designed. 
A dedicated bench test system evaluated these bases and investigated saturation and suppression responses in Hamamatsu R12699 PMTs. 
The four-capacitor base (BASE-2) meets PandaX-xT requirements for PMTs at different gains from 2.7 to $5.5\times10^6$, and its saturation and suppression effects were tested and characterized.
For a deeper understanding of the base response, an LTSpice-based circuit simulation model was developed; this model reproduces saturated/suppressed waveforms, captures the key physics of the responses, and facilitates base circuit design and evaluation. 
Finally, the combined simulation and bench test approach guides base design and optimization, improving detector performance and supporting future saturation/suppression correction for PandaX-xT's data analysis.

Future work will focus on: (1) optimizing desaturation capacitor materials to reduce intrinsic radioactivity; (2) integrating the saturation/suppression correction model into the data analysis chain; (3) extending the simulation to other PMT models for cross-validation.

\section*{Acknowledgment}
This project is supported in part by grants from the National Key R\&D Program of China (Nos. 2023YFA1606200, 2023YFA160620[1,2,3,4, 
]), National Science Foundation of China (Nos. 12090060, 12505132, 1209006[2], U23B2070), and by Office of Science and Technology, Shanghai Municipal Government (grant Nos. 21TQ1400218, 22JC1410100, 23JC1410200, ZJ2023-ZD-003), and the Natural Science Foundation of
Shanghai (No. 24ZR1437100). We thank for the support by the Fundamental Research Funds for the Central Universities. We also thank the sponsorship from the Chinese Academy of Sciences Center for Excellence in Particle Physics (CCEPP), Thomas and Linda Lau Family Foundation, New Cornerstone Science Foundation, Tencent Foundation in China, and Yangyang Development Fund. Finally, we thank the CJPL administration and the Yalong River Hydropower Development Company Ltd. for indispensable logistical support and other help.

\end{document}